\documentclass[%
reprint,
superscriptaddress,
 amsmath,amssymb,
 aps,
 pra,
]{revtex4-1}

\usepackage{graphicx}
\usepackage{dcolumn}
\usepackage{bm}

\usepackage{todonotes}
\usepackage{csquotes}
\usepackage{balance}

\usepackage[utf8]{inputenc}

\usepackage{titlesec}
\titlespacing*{\section}{0pt}{1.1\baselineskip}{\baselineskip}
\titlespacing*{\subsection}{0pt}{1.1\baselineskip}{\baselineskip}

\newtheorem{problem}{Problem}

\begin{document}

\title{NISQ circuit compilation is the travelling salesman problem on a torus}

\author{Alexandru Paler}
\email{alexandrupaler@gmail.com}
\affiliation{%
    Institute of Integrated Circuits, Johannes Kepler University
}
\affiliation{%
    Transilvania University of Brașov, Romania
}
 
\author{Alwin Zulehner}
\author{Robert Wille}
\affiliation{%
    Institute of Integrated Circuits, Johannes Kepler University
}

\begin{abstract}
Noisy, intermediate-scale quantum (NISQ) computers are expected to execute quantum circuits of up to a few hundred qubits. The circuits have to conform to NISQ architectural constraints regarding qubit allocation and the execution of multi-qubit gates. Quantum circuit compilation (QCC) takes a nonconforming circuit and outputs a compatible circuit. Can classical optimisation methods be used for QCC? Compilation is a known combinatorial problem shown to be solvable by two types of operations: 1) qubit allocation, and 2) gate scheduling. We show informally that the two operations form a discrete ring. The search landscape of QCC is a two dimensional discrete torus where vertices represent configurations of how circuit qubits are allocated to NISQ registers. Torus edges are weighted by the cost of scheduling circuit gates. The novelty of our approach uses the fact that a circuit's gate list is circular: compilation can start from any gate as long as all the gates will be processed, and the compiled circuit has the correct gate order. Our work bridges a theoretical and practical gap between classical circuit design automation and the emerging field of quantum circuit optimisation.
\end{abstract}

\maketitle

\section{Introduction}
\label{sec:intro}

The first general purpose quantum computers, which are called noisy, intermediate-scale quantum (NISQ) computers \cite{preskill2018quantum}, operate on a few hundred qubits and do not support computational fault-tolerance. The IBM Q Experience computers, which fall into the NISQ category, have sparked the interest in the automated compilation of arbitrary quantum circuits. Near-term applications of NISQ may be used to explore many-particle quantum systems or optimisation problems, and the executed circuits are not expected to include sequences longer than 100 gates\cite{preskill2018quantum}. Although this is a serious limitation, it is hoped that hardware quality will increase such that longer circuits may be executed.

NISQ compilation is motivated in part by the different architectures, but more important by the technical limitations of NISQ hardware, such as qubit and quantum gate fault rates, gate execution time etc. Before executing a quantum computation the corresponding circuit has to be adapted for the particularities of the NISQ computer.

\subsection{Background}

\begin{figure}[t!]
\centering
\includegraphics[width=0.7\columnwidth]{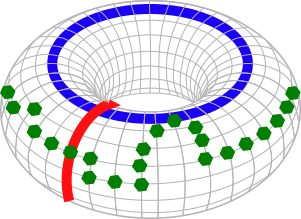}
\caption{The search landscape of QCC is a discrete two dimensional torus. The blue circle represents the timeline of the quantum circuit being compiled. The green vertices (only a few are drawn) are qubit allocation configurations encountered during compilation. The red circle is connecting all the qubit allocations configurations available while compiling one of the circuit's gate. QCC compilation forms a loop formed of green vertices. The solution loop intersects non-trivially all possible red loops on the torus.}
\label{fig:torus1}
\end{figure}

In order to show that the problem of quantum circuit compilation is equivalent to a travelling salesman on a torus (e.g. Fig.~\ref{fig:torus1}) we introduce the following background material. 

\begin{figure}[h]
\centering
\includegraphics[width=0.9\columnwidth]{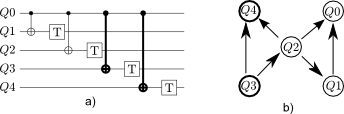}
\caption{a) Quantum circuit example; b) Example of a coupling graph where the vertices $Q3$ and $Q4$ are not adjacent to $Q0$. Considering the graph from b), the CNOTs $Q0 \rightarrow Q3$ and $Q0 \rightarrow Q4$ are remote. The remote CNOTs are highlighted in a).}
\label{fig:remotecnot}
\end{figure}

We will use NISQ, quantum computer, and chip interchangeably. For the purpose of this work, a chip is described entirely by the set of hardware qubits, also called registers \cite{siraichi2018qubit}, and the set of supported interactions. The computer is abstracted by a \emph{coupling graph} (e.g. Fig.~\ref{fig:remotecnot}b), where the registers are the vertices, and the edges are the supported multiqubit gates between vertex tuples. In a directed coupling graph $G=(V,E)$, having $|V|=q$ and $|E| \leq q(q-1)$, the edges stand for the CNOTs supported between pairs of physical qubits. The edge directions indicate which qubit is control or target. If the computer supports both CNOT directions between a pair of qubits, there are two directed edges between the corresponding graph vertex pairs. Current NISQ devices do not restrict CNOT direction, and $G$ graphs are nowadays mostly undirected.

In general, NISQs do not have all-to-all connectivity between the registers, and do no support the arbitrary application of multiqubit gates. Consequently, not all the CNOTs of a circuit can be executed without further adjustment.

The quantum circuit compilation (QCC) problem is: \emph{for a given coupling graph and a quantum circuit $C$, compile a circuit $C'$ which is functionally equivalent to $C$ and compatible with the coupling graph}.

We use the following operations to solve QCC: 1) qubit allocation; 2) CNOT gate scheduling, and 3) circuit traversal - choosing the order in which the CNOTs are compiled. The first two are practically already methodological parts of established quantum circuit design frameworks such as Cirq and Qiskit. A theoretical analysis of the first two was provided in \cite{siraichi2019qubit}. The third operation is based on circular CNOT circuits as introduced in \cite{paler2016circular}.

Each of the three operations can be attached to an optimisation problem. Each of those problems is directly connected to the execution of a \textbf{remote CNOT}, which is defined as the gate that has to be executed between circuit qubits allocated on non-adjacent NISQ registers (e.g. Fig.~\ref{fig:remotecnot}). The compilation of a remote CNOT introduces additional gates into $C'$, because the qubits have to be effectively moved across the chip until these are on adjacent registers. A correctly compiled $C'$ has no remote CNOTs. It is assumed that the NISQ chip has at least as many registers as $C'$.

\begin{figure}
\centering
\includegraphics[width=0.9\columnwidth]{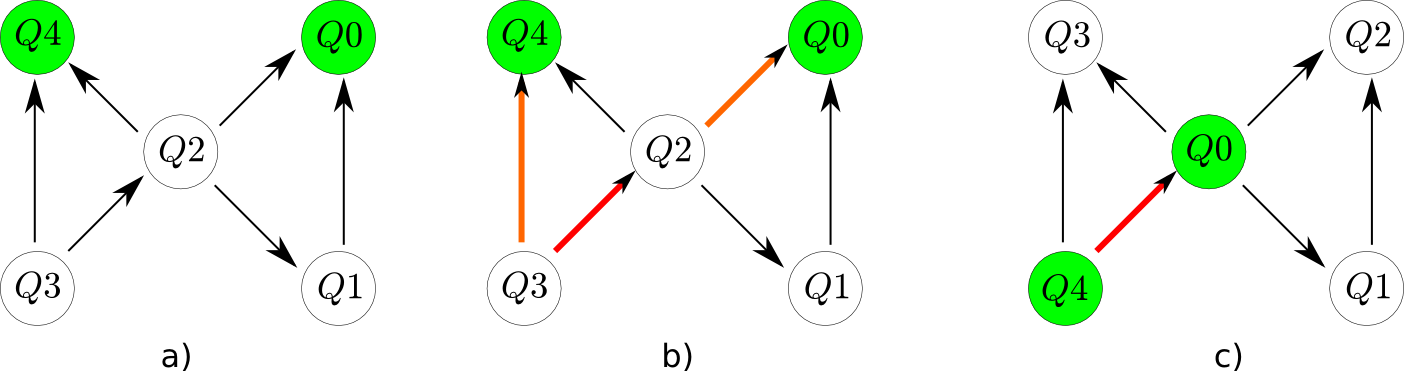}
\caption{Finding the edge where to execute a remote CNOT: a) The CNOT $Q0 \rightarrow Q4$ needs to be implemented, but the qubits are not adjacent. b) Depending on a given cost function, it is determined that moving the qubits to the endpoints of the red edge (between $Q3$ and $Q2$) is the most cost effective method to achieve $Q0 \rightarrow Q4$. c) A new qubit allocation configuration is generated after the $Q_0$ and $Q_4$ were moved across the chip.}
\label{fig:heuristic}
\end{figure}

We define the qubit allocation problem with respect to the effect of compiling remote CNOTs. Fig.~\ref{fig:heuristic} includes two allocation configurations. The labels inside the coupling graph vertices represent the allocated circuit qubits. After moving $Q_4$ from one register to another, the configuration from Fig.~\ref{fig:heuristic}a changes to the one from Fig.~\ref{fig:heuristic}c.

\begin{problem}
\label{pb:1}
\textbf{Qubit allocation}: Assign circuit qubits to NISQ registers, such that the compiled $C'$ has a minimal cost.
\end{problem}

The cost mentioned in the Problem 1 could be gate count, circuit depth etc. For example, the cost can be expressed in terms of physical CNOT gates and assuming a linear nearest neighbour architecture, in Fig.2 the cost of implementing the first CNOT is zero, the second remote CNOT has a cost of six because two SWAP gates may be necessary etc.

For the purpose of this work, the compilation of remote CNOTs to the NISQ chip is a kind of gate scheduling (see Problem~\ref{pb:2}) and we present an example in Fig.~\ref{fig:heuristic}. Automatic approaches for gate scheduling range from global reordering of quantum wires \cite{wille2014exact} to application of circuit rewrite rules \cite{saeedi2011synthesis}. Gate scheduling has been performed even manually, by designing circuits that conform to the architectural constraints \cite{fowler2004implementation, boixo2018characterizing}.

\begin{problem}
\label{pb:2}
\textbf{Gate scheduling}: Choose the coupling graph edge where to execute a remote CNOT.
\end{problem}

Gate scheduling is a sub-problem of the qubit allocation problem, because scheduling is performed once qubits are allocated. However, for an exact solution, finding the best initial allocation requires iterating through all possibilities which in turns implies that gate scheduling has to be calculated each time. From this perspective, QCC is at least as complex as the qubit allocation problem.

Regarding scheduling, it is not obligatory to start compiling from the first remote CNOT of $C$. One can start from an arbitrary gate, as long as the resulting $C'$ will respect the original order from $C$. For example, if $C$ consists of three remote CNOTs $g_1$, $g_2$ and $g_3$, the compilation could start with $g_2$, followed by $g_3$ and finally $g_1$. However, $C'$ would need to execute the correct order of compiled gates $g_1', g_2', g_3'$.

\begin{problem}
\label{pb:3}
\textbf{Circuit traversal}: Determine the order in which the gates of $C$ should be compiled, such that the cost of $C'$ is minimised. The chosen order has to be a valid topological sorting of $C$.
\end{problem}

\subsection{Related work}
\label{sec:related}

Most NISQ devices have a topology which is not compatible with the quantum circuits that have to be executed on them. Those circuits need to be accordingly modified. Originally, this has been done by adapting the quantum circuit in a systematic manner (e.g.~\cite{fowler2004implementation}). However, such an approach obviously is not feasible---particularly with increasing size of the considered quantum circuit. 

In the past, a huge variety of of methods addressing this problem have been proposed. While some of them (e.g.~\cite{wille2014exact,DBLP:journals/corr/RahmanD15,DBLP:conf/dac/WilleBZ19,DBLP:conf/aspdac/ItokoRIMC19}) aim to solve QCC in an exact fashion (i.e.~generating minimal solutions) most of them provide heuristics. Heuristics are much more established solutions for QCC, while exact approaches are mainly used for evaluation purposes (i.e.~checking how far heuristics are from the optimum) or to generate quantum circuits for certain ``building block''-functionality. 

Most of the available heuristics employ a \emph{swapping}-scheme, i.e.~they insert remote CNOT and SWAP operations into the originally given quantum circuit that exchange the state of two physical qubits whenever they do not satisfy a connectivity constraint. By this, the mapping of the logical qubits of the quantum circuit to the physical ones of the hardware changes dynamically, i.e.,~the logical qubits are moved around on the physical ones. Approaches following this scheme include e.g.~\cite{DBLP:conf/aspdac/ItokoRIMC19,saeedi2011synthesis,wille2016look,zulehner2017efficient,8702439,DBLP:conf/rc/HattoriY18,DBLP:conf/rc/MatsuoY19}.

Other approaches use a \emph{bridging}-scheme which 
does not dynamically change the mapping of the logical qubits to the physical ones: CNOT gates that violate the connection constraint are decomposed into several CNOT gates that bridge the \enquote{gap}. This scheme has the advantage that, given the initial mapping, determining the mapped circuit is straightforward. On the other side, it often leads to more costly solutions since the number of CNOT operations required to realize bridge gates grows exponentially. Approaches following this scheme include e.g.~\cite{DBLP:journals/corr/RahmanD15,dueck2018optimization,de2019cnot,DBLP:conf/aspdac/ItokoRIMC19,DBLP:journals/integration/ItokoRIM20}.

\subsection{Complexity of QCC}

Multiple approaches to showing the complexity of QCC have been presented. One of the first, Maslov\cite{maslov2008quantum} demonstrated that a variant of QCC is NP-complete by showing that it implies the search of a Hamiltonian cycle in a graph. In the context of our QCC formulation, the work of \cite{maslov2008quantum} is concerned with optimal solutions to the qubit allocation problem when the blue torus edge weights are determined by the physical gate execution times along the longest input-output gate chain.

QCC has also been considered a search problem according to \cite{siraichi2019qubit}, which includes a detailed review of the methods used for determining the complexity class. It has been recently discussed that the complexity of QCC optimisation is NP-hard \cite{nash2020quantum} by comparing QCC with the optimisation of fault-tolerant quantum circuits protected by the surface code \cite{herr2017optimization}.

The authors of  \cite{botea2018complexity} have shown that QCC as a discrete optimisation of a circuit's makespan is NP-complete for QAOA circuits. The proof from \cite{botea2018complexity} on a reduction from the Boolean satisfiability problem (SAT) to the QCC problem, and was applied to circuits consisting of two types of two-qubit gates: SWAP and PS (phase separation). The proof did not rely on any particular ordering of the gates in the circuit.  Such circuits can be decomposed with a constant overhead into the circuits we consider in this work (CNOT gates and single qubit gates). PS gates can be decomposed using the KAK decomposition \cite{tucci2005introduction, zulehner2019compiling} into CNOTs and single qubit gates, and the SWAP gate can be decomposed into three CNOTs.

Moreover, a method for optimising QAOA circuits by taking commutativity into account was presented by \cite{tan2020optimal}. Therein the authors show very convincingly that theorem proving (e.g. Z3 solver) and SAT solvers do not scale for practically large compilation problems (more than 100 qubits and deep circuits): the search space of QCC as an NP-complete problem is still exponential even when the number of variables is reduced exponentially. The work of \cite{tan2020optimal} combined with the theoretical approach from \cite{botea2018complexity} highlight the importance of QCC heuristics.

QCC has been presented as an application of temporal planning \cite{venturelli2018compiling}, too. In general, temporal planning can have a higher complexity than NP-complete. For example, concurrent temporal planning is EXPSPACE-complete \cite{rintanen2007complexity}. This, however, does not imply that QCC would be EXPSPACE-complete. In fact, in domain-independent planning, it is not uncommon that a planning system attacks a domain with a lower complexity that the complexity of the AI planning variant that the planning system at hand can handle. This is done for the convenience of using a readily available off-the-shelf system, when a domain-specific solver is not necessarily available.

\section{Methods}
\label{sec:methods}

We present the construction of how QCC can be solved as the travelling salesman problem. To this end, we illustrate the construction of the QCC torus.

\subsection{Arranging qubit allocations in a circle}
\label{sec:perm}

Allocating circuit qubits to NISQ registers can be expressed as a permutation vector of length $q$. For example, assume that $Q_i$ are the qubits of a circuit $C$, and $H_i$ are the registers of a computer, for $i \leq q$. Both the computer and the circuit have $q=5$ qubits. The permutation $p_1=(0,1,2,3,4)$ is the trivial allocation where $Q_i$ are allocated to $p_1[Q_i]=H_i$: circuit qubit $Q_0$ at register $H_0$, qubit $Q_1$ at register $H_1$ etc. Another example is the permutation $p_2=(2,1,0,4,3)$, where $Q_0$ is allocated to $H_2$, $Q_1$ is allocated to $H_1$ etc.

In the following, a \textbf{configuration} is a permutation that represents how circuit qubits are allocated to NISQ registers. The terms permutation and configuration will be used interchangeably. For example, $p_1$ and $p_2$ are configurations, too.

The set of all permutations forms a symmetric group with $q!$ elements. The group has $q-1$ transposition generators. A transposition swaps two elements of the permutation, and keeps all other entries unchanged. Any group element is expressed through a non-unique sequence of transposition generators.

\begin{figure}[h!]
\centering
\includegraphics[width=0.4\columnwidth]{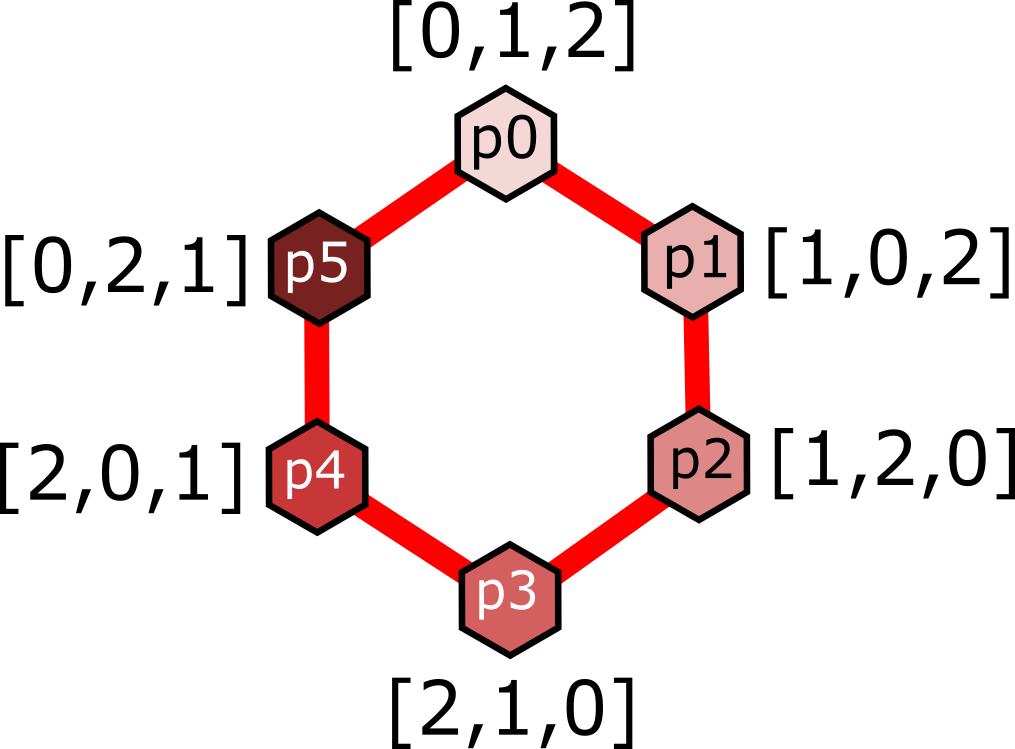}
\caption{Circular arrangement of $6=3!$ configurations. In order to avoid confusion with a coupling graph, vertices are polygons. The permutation vector associated to each vertex is placed next to the vertex. For example, $[0, 1, 2]$ is next to $p0$. The edge represents transpositions. For example, the permutation $p0$ is transformed into $p1$ by transposing $0$ and $1$. The edges and vertices are red in order highlight the correspondence to Fig.~\ref{fig:torus1}. The weights along the red circles are zero (Sec.~\ref{sec:weights}).}
\label{fig:excyc1}
\end{figure}

The group structure can be visualised as a graph. The elements are vertices, and edges are transpositions connecting the vertices. If all group elements are exhaustively enumerated, the graph is a circle (e.g. Fig.~\ref{fig:excyc1}) with $q!$ vertices. There exist more compact representations of the group, such as the complete graph $K_q$. Without affecting the generality, the exhaustive representation is preferred in this work.

\subsection{The circuit as a circle of CNOTs}
\label{sec:chain}

The compilation problem has been reduced to scheduling the execution of CNOTs, remote or not. Quantum circuits are often manipulated as directed acyclic graphs (DAGs) with vertices for quantum gates. Edge directions reflect the gate ordering inside the circuit. For the purpose of this work, the DAG representation is replaced by the equivalent (blue) circle of CNOTs \cite{paler2016circular}. The order of the vertices on the blue circle encodes one of the equivalent topological orderings from the DAG. In general, gate commutativity may be used to improve the compiled circuit (see Appendix on the backtracking method). In particular, all equivalent DAG topological orderings may need to be considered. The latter is equivalent to commuting gates from the chosen topological ordering with an identity gate.

\begin{figure}
\centering
\includegraphics[width=0.9\columnwidth]{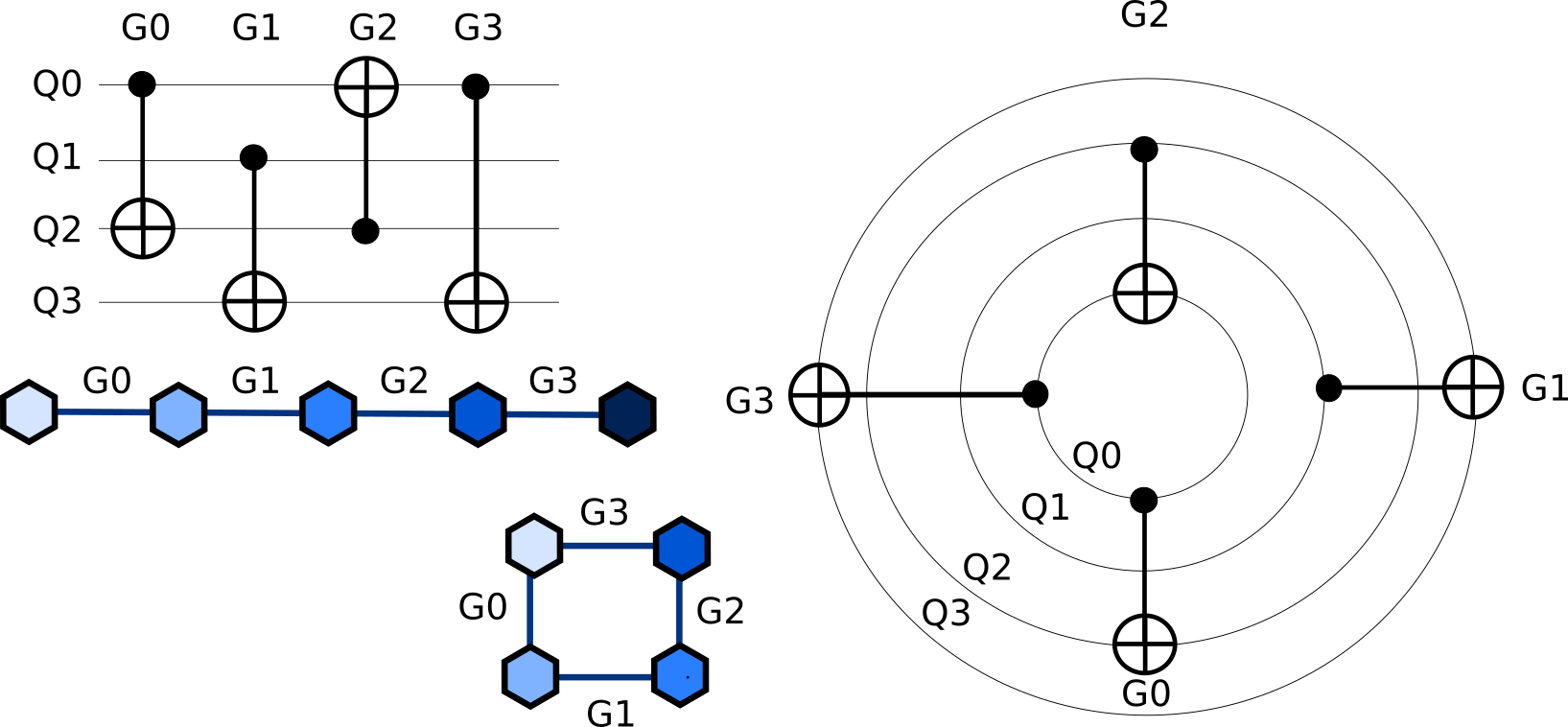}
\caption{Example: A quantum circuit with four CNOTs named $G_0\ldots G_3$ is drawn as a circular CNOT circuit having the wires as concentric circles. The normal circuit is used to obtain a chain of configurations after the application of each CNOT. The edges in the chain represent the CNOT. The ends of the chain can be joined to form a circle. The edges and vertices are blue in order to highlight the correspondence to Fig.~\ref{fig:torus1}}.
\label{fig:exchain}
\end{figure}

A circle is obtained as follows: a) only the CNOTs are kept from the circuit, and other gates are discarded (e.g. the T gates from Fig.~\ref{fig:excyc1}), b) the wire endpoints corresponding to input and output are joined together. Fig.~\ref{fig:exchain} is an example of obtaining a circular CNOT circuit. Pairs of adjacent vertices in the chain represent the qubit allocation configurations before and after a remote CNOT was compiled. 

It is possible to start compiling a circuit $C$ from any gate $g$ and not necessarily from the first gate. The circular CNOT circuit supports the correctness of this observation. Let us consider that the circuit $C$ is the application of a sequence of two sub-circuits $A$ and $B$, such that $C=AB$. Moreover, we model QCC as a function that computes $QCC(C)=C'$, where 
$QCC(C)=QCC(A)QCC(B)=C'$. 

Instead of starting with the first gate of $A$, we assume that compilation starts from sub-circuit $B$ and that the CNOT circle is traversed in the correct order. Along the circular traversal $A^\dagger$ will be compiled instead of $A$. The compilation result will be circuit $D$ for which $D=QCC(B)QCC(A^\dagger)$.

However, it is possible to to reconstruct $C'$ by inverting the gate list of $QCC(A^\dagger)$ such that  $QCC(A^\dagger)^\dagger=QCC(A)$. This divide and conquer approach does not imply that a greedy approach can solve QCC efficiently. It still is a combinatorialy difficult to choose the \emph{best} gate from $C$ to start compilation from.

Starting the traversal of CNOT circles from arbitrary positions can be advantageous for reducing the total cost of the compiled circuit.

It is not guaranteed that $Cost(C,D)=Cost(C,C')$, such that a heuristic approach to QCC could be to start compiling from different gates of $C$.

\subsection{Unfolding the torus}

\begin{figure}[t!]
\centering
\includegraphics[width=0.8\columnwidth]{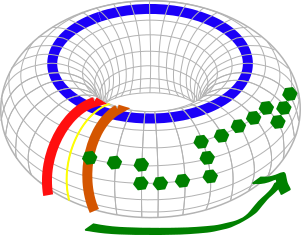}
\caption{After cutting the torus along the yellow line, the start circle is brown and the stop circle is red. Compilation is the problem of connecting the circles by a loop of green vertices, such that the sum of the edge costs is minimal.}
\label{fig:torus2}
\end{figure}

\begin{figure}[h]
\centering
\includegraphics[width=0.8\columnwidth]{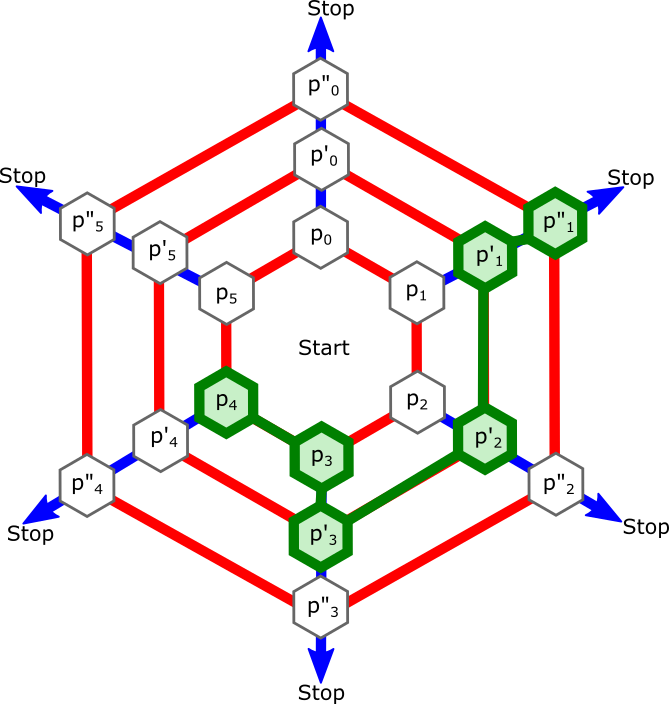}
\caption{Example of a search diagram with $3!$ configurations. The red configuration circles are concentric. The red circles are interconnected by  CNOT circles that were cut like in Fig.~\ref{fig:torus2}. Search starts in the centre of the diagram and stops at one of the endpoints of the radial CNOT circles. The green vertices indicate the path found by the compiler.}
\label{fig:rad1}
\end{figure}

The circular graph of configurations and the CNOT circle can be combined to a \emph{torus} (e.g. Fig.~\ref{fig:torus1},~\ref{fig:torus2}). The torus has a discrete structure, which can be used to visualise and analyse QCC. For visualisation purposes, the torus can be cut and unfolded to a planar structure. We will resort to a single cut along the configuration circle. The result will be a two dimensional diagram like the one in Fig.~\ref{fig:rad1}. Let one side of the cut be called the start-circle and the other side the stop-circle.

\begin{figure}[h]
\centering
\includegraphics[width=0.5\columnwidth]{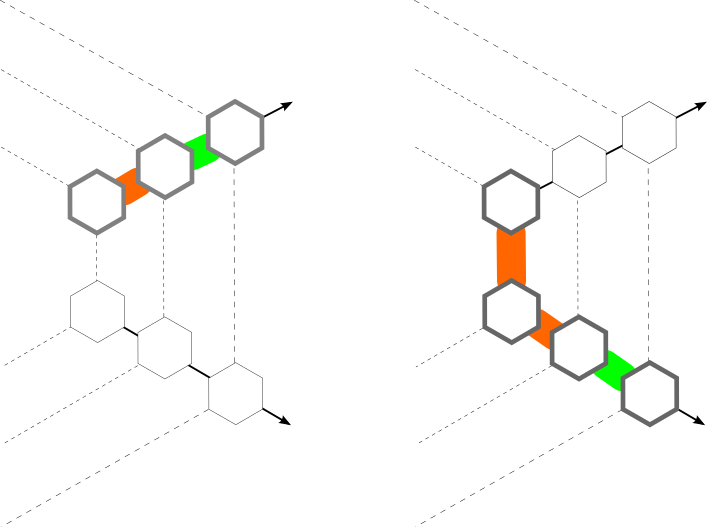}
\caption{Compilation visualised as a movement on torus. Two CNOTs are compiled (orange and green). left) The compilation of the two CNOTs is a movement along the CNOT circle. right) The compilation of the first CNOT includes first a movement along a configurations circle. Both compilations (left and right) result in a correct circuit, but the compiled results have different costs.}
\label{fig:sdmoves}
\end{figure}

As shown in Fig.~\ref{fig:rad1} and Fig.~\ref{fig:sdmoves}, a hypothetical quantum compiler will traverse vertices of the torus. The number of torus vertices is the total number of states the compiler should consider, and there are $q! \times |C|$ states. By restating Problem~\ref{pb:1}, the compiler will find a path from the start circle to the stop circle (Fig.~\ref{fig:rad1} and Fig.~\ref{fig:torus2}). There is a combinatorial number of paths of various lengths between pairs of start-stop vertices. QCC executes, in the best case, linearly in the number of circles traversed between start and stop.

\subsection{Edge weights}
\label{sec:weights}

The edges connecting the vertices of the torus are weighted. Two extreme cases are possible: a) all edges have weight zero; b) all edges have equal weight. The first case is not realistic in the context of QCC. The second case arises when the NISQ device has all-to-all connectivity, such that the shortest path between a start and a stop circle is given by the straight traversal of a CNOT circle. For the purpose of this work, the red edges (configuration edges) have zero weight, and the blue edges (CNOT edges) have non-zero weight. The motivation for this model is twofold.

First, our goal is to show that QCC is TSP (see Section~\ref{sec:nphard}), and we have chosen the generalised TSP form as presented in \cite{noon1993efficient}. This TSP form uses the concept of connected city clusters. The movement within a cluster has cost zero, but the movement between clusters non-zero. In our case, the red rings are the clusters and the blue rings are the connections between the clusters.

Second, instead of weighing the red edges, we consider their cost as part of the blue edge weights. Each blue edge traversal requires compiling a (remote) CNOT. The compilation is thus determined by: a) the cost of implementing the transposition resulting by moving along the red ring (a new start qubit allocation configuration from which the CNOT is compiled), and b) the cost of effectively scheduling the remote CNOT.

Additionally, we note that by joining the first and last red rings the qubit allocation configuration has to be the same, in general. This is the case, when compilation does not start from the first gate of the circuit (cf. Section~\ref{sec:chain}) and needs to reconstruct the solution. After reconstructing the solution, however, the wire permutations before the first and last gate can be removed -- these are simple wire relabelling operations. As a result, configuration changes on the start/stop circle come for free and are not considered in the compilation cost, because no gates need to be inserted in the circuit. For example, in Fig.~\ref{fig:sdmoves} the orange traversal of the configuration ring has cost zero.

This brings us to the particular QCC scenario, which we assume being the common one, when compilation starts from the start ring (e.g. brown in Fig.~\ref{fig:torus2} corresponds to the first gate from the uncompiled circuit) and ends on a different vertex of the same ring. Different vertices on the same ring refer to different qubit allocation configurations. Therefore, in particular, it is an acceptable solution to end on the same ring, but on a different vertex.

We mentioned that the weights may be, for example, the number of physical CNOTs necessary to implement a remote CNOT. In general, edge weights are assigned by a \textbf{cost function}. It is the task of the cost function to extract information from the circuits and the coupling graph. It is the task of the cost function, for example, to perform topological analysis of the circuit and coupling graph \cite{ferrari2018demonstration}. The cost of gate compilation could include also lookahead information, similarly to how this was performed for example for linear nearest neighbour architectures \cite{wille2016look}.

Formulating explicit cost functions does not fall within the scope of this work. As shown in \cite{tan2020optimal}, once the cost functions are specified, formulating the optimisation objective is a highly nontrivial task. The optimisation objective is for exact QCC methods like the actual code implementation is to the heuristic QCC methods. Therefore, even if we would specify the exact functions, the optimality of the compiled circuit would depend on the time-space trade off allowed by the heuristic implementation. In particular, just as examples: a) if the optimisation goal is the minimum number of SWAPs one could use the MI strategy from the Appendix; b) for minimising depth, and by making no assumptions about gate execution time like in \cite{maslov2008quantum}, the optimisation goal would be makespan  \cite{botea2018complexity}.

From the perspective of an arbitrary function $Cost(C, C')$ that calculates the cost of compiling $C$ into $C'$ (similar to discussion in (Sec.~\ref{sec:nphard}), we can state that QCC optimisation is to find a circuit $M$ such that $Cost(C, M) = min(Cost(C, C_l))$, where $C_l$ is a loop on the torus. The best $C_l$ loop has the minimum sum of the traversed edges.

\section{Results}

The landscape of QCC is a discrete torus obtained from the Cartesian product of two circles. One of the circles refers to the group structure of the qubit allocations possible when scheduling a gate (ie. the red circle in Fig.~\ref{fig:torus1}). The other circle is generated by the fact that the CNOTs of a circuit can be arranged in a circular form (ie. the blue circle in Fig.~\ref{fig:torus1}) \cite{paler2016circular}.

The torus includes $|C|$ red circles - one for each gate from $C$. There are $q!$ blue circles: one for each possible permutation of circuit qubits to NISQ registers. The details of constructing the torus were presented in Sec.~\ref{sec:methods}.

\subsection{QCC is a TSP}
\label{sec:nphard}

In the following we show that QCC is a travelling salesman problem (TSP). The Appendix includes a backtracking formulation of QCC as TSP. Independent of this work, the authors of \cite{siraichi2019qubit} have decomposed the compilation problem into two steps: qubit allocation and scheduling of multi-qubit gates. In practice, this approach has already been followed by quantum circuit frameworks such as Cirq and Qiskit: the circuit qubits are mapped to the NISQ device, and then the circuit gates are scheduled. For QCC benchmarking purposes, the two-step approach has also been used by \cite{tan2020optimality}. We augment the QCC decomposition by including the circular CNOT structure. This will be useful for analysing the problem complexity. The exact complexity depends on how the cost function is implemented and evaluated.

We use the following definitions:
\begin{itemize}
    \item \textbf{A solution} is any loop that intersects non-trivially the red circle from Fig.~\ref{fig:torus1}. There is a combinatorial number of potential solutions.

    \item \textbf{The minimum solution} is the loop for which the sum of the edge weights is minimal (Sec.~\ref{sec:weights}).
\end{itemize}

According to the discussion in Sec.~\ref{sec:weights}, only the edges along the CNOT circles have non-zero weights. Each solution is the sum of $|C|$ weights $Cost_l = \sum_0^{|C|}w_{p,q}$, where $l$ is the index of the solution and $p,q$ are the indices of the configurations connected by the edge that has weight $w_{p,q}$. The solution of QCC is $min(Cost_l)$, for $l \leq (q!)^{|C|}$ when the exhaustive enumeration of the configurations is used. In the light of the definitions of Problems~\ref{pb:1}-\ref{pb:3}, where a minimum cost circuit is searched, QCC is an example of combinatorial optimisation. 

We show that QCC is a generalised TSP (GTSP). The original TSP problem is defined for a number of cities, for which the distances between pairwise cities are known. TSP answers the question: what is the shortest possible route visiting all the cities and returning to the origin city? In GTSP the cities are arranged into clusters, and the edges connecting the cities inside the cluster have weight zero \cite{noon1993efficient}. At least one city from each cluster has to be visited on the shortest path \cite{noon1993efficient}.

QCC is GTSP when considering each red configuration ring of the $|C|$ as a cluster of cities. Moreover, the zero weight cluster edges are consistent to how the weights along the configuration rings are set in Sec.~\ref{sec:weights}. There are $|C|$ configurations circles in the torus. The distances between the cities are the weights along the CNOT edges. The salesman is expected to traverse at least once each configuration circle between the red start circle and the brown circle from Fig.~\ref{fig:torus2}.

The fact that the configuration rings are arranged in a circle does not make the problem easier. Assuming that $|C|$ has only three remote CNOTs, then there are only three clusters for which the GTSP has to be computed. However, the arrangement of the three clusters corresponds to a complete graph $K_3$ -- the smallest instance of GTSP. Increasing the length of the circuit increases the number of clusters, but does not reduce the complexity of the optimisation problem.

The decision GTSP version of QCC answers the question: is there a route/loop of cost less than a specified $Cost_{route}$? Any potential solution can be verified by tracking the proposed solution loop along the torus. Because of its complexity, QCC has to be solved using heuristics. Benchmarking circuits for which the minimum $Cost_{route}$ is known beforehand \cite{tan2020optimality} are a good way to evaluate the performance of the heuristics.

\subsection{QCC is a ring}

The discrete torus shows that QCC, from the perspective of discrete mathematics, is a ring with the two QCC-operations being: 1) qubit allocation; 2) gate scheduling.

We can define commutativity in a manner compatible with quantum circuit execution. Very informally, two QCC-operations are commutative iff the computation implemented by the circuit is unchanged after reordering the QCC-operations. Consequently, qubit allocation is commutative because it is effectively a renaming of wires. It does not matter in which order the qubits are allocated, this does not change the computation. QCC-gate scheduling is not commutative because, in general, two CNOTs are not commutative. Consequently, the circle of allocations is the illustration of an Abelian group, and the CNOT-circle represents a monoid. The Abelian group and the monoid form the discrete torus where traversal are unidirectional. We leave a formalisation of the mathematical structure of QCC for future work.

\section{Discussion and Conclusion}
\label{sec:concl}

NISQ compilation is receiving increased attention, due to its practical industrial relevance. In this work, the QCC problem was decomposed into a set of sub-problems, whose individual solution is found by traversing circles. This enabled the formulation of QCC as a travelling salesman problem along a torus. We have implemented the TSP approach to QCC compilation at \url{https://github.com/alexandrupaler/k7m} and we have used it as part of a machine learning approach to QCC in \cite{paler2020machine}. 

The torus structure presented has the potential to generate other efficient heuristics for the QCC compilation. Exact QCC methods \cite{wille2014exact, tan2020optimal} scale poorly, because these are as fast as the underlying solver. The highly regular and cyclic structure of the torus search space may inspire improved variable encodings such that exact layout methods can be pushed in the area of 100-qubit circuits.

The TSP formulation hints at the conceptual similarities between QCC and the automatic design of quantum optical experiments \cite{krenn2020computer}. The latter consist of discrete optical elements, which can be placed in a combinatorial (mapping steps) number of experimental configurations formed by different devices (scheduling step). At the same time, forming loops on the torus shows that QCC is similar to a dynamic optimisation problem \cite{cruz2011optimization}, and that it would be reasonable to expect methods based on ant colonies or evolutionary algorithms for solving QCC.

Finally, because quantum methods have been applied to TSP (using QAOA in \cite{radzihovsky2019qaoa}, using Grover's algorithm by \cite{srinivasan2018efficient}), our work opens the possibility to optimise quantum circuits with quantum computers.

\section*{Acknowledgement}

We are very grateful to Adi Botea for his technical input, feedback and suggestions. We acknowledge the input of Razvan Andonie on optimisation problem complexity classes, as well as the technical corrections proposed by Daniel Herr. This work was funded by the Linz Institute of Technology project CHARON, the Google Faculty Research Award FRA\textunderscore ANGELICO, and the NUQAT project of Universitatea Transilvania Brasov.

\bibliographystyle{plain}
\bibliography{bibi}

\section*{Appendix}

\subsection{Moving on a (red) configurations circle}

\begin{figure}[h!]
\centering
\includegraphics[width=0.9\columnwidth]{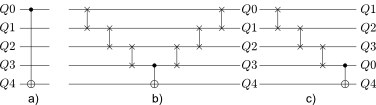}
\caption{Two strategies of inserting SWAPS: a) initial circuit; b) Move-Interact-Move (MIM); c) Move-Interact (MI). The initial configuration is maintained after applying MIM (e.g. $(0,1,2,3,4)$). The MI strategies results in the $(1,2,3,0,4)$ configuration.}
\label{fig:swaps}
\end{figure}

At least two strategies are possible for compiling \emph{a single remote CNOT}. The methods are illustrated in Fig.~\ref{fig:swaps}. The first strategy is MIM (abrv. for move-interact-move): move one of the qubit states on a wire next to the other qubit's wire, interact the qubits, and then swap back to the original wire. The second strategy is called MI (abrv. for move-interact) and is similar to the first one but without swapping back the moved qubit state.

Applying MIM once introduces $2d$ SWAP gates in the circuit, while the MI strategy only $d$ SWAPS, where $d$ is the distance between the remote wires. A straightforward distance function could be, for example, the Manhattan distance which can be used for LNN as well as grid NISQ architectures. For a given permutation $p$, after applying MIM, the resulting permutation is also $p$. 

On the contrary, after an MI swap, the resulting permutation is a $p'$, obtained through the sequence of transpositions representing the SWAP gates. Although MI introduces less SWAPS, it increases the complexity of the compilation problem: each remote CNOT will result in a new permutation, such that the circuit qubit allocation configuration is \emph{evolving} after each CNOT.

In the presence of evolving configurations, state of the art compilation methods are solving the following problem: find an optimal circuit consisting entirely of SWAP gates that transforms a current permutation $p_{in}$ to a permutation $p_{out}$ such that a given batch of remote CNOTs can be implemented on the given coupling graph. In other words, an optimal sequence of transpositions is sought, such that $p_{out}$ conforms to a set of constraints imposed by all the CNOTs to implement. During the search of a SWAP circuit, or after a SWAP circuit was found, it is checked that $p_{out}$ conforms to the coupling graph.

\begin{figure}[h]
\centering
\includegraphics[width=0.9\columnwidth]{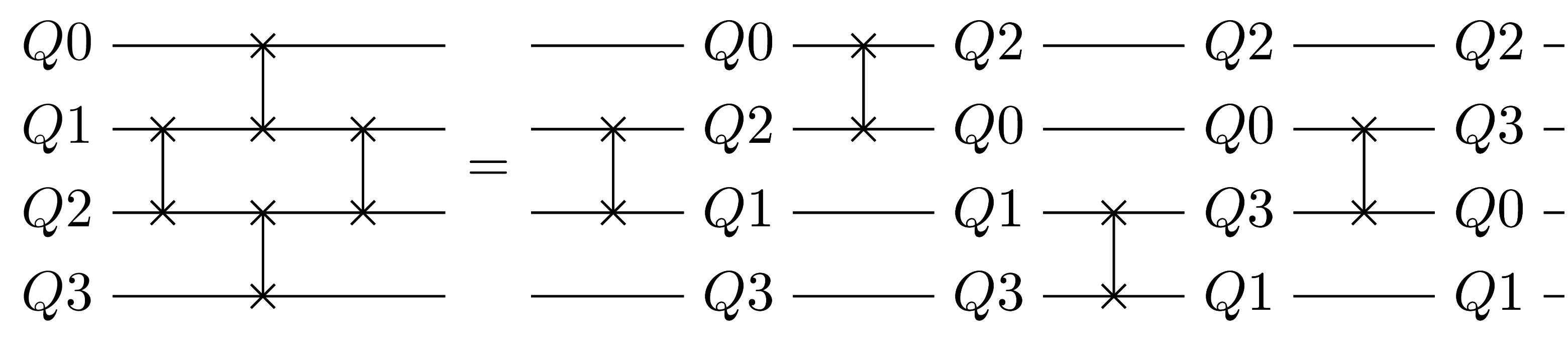}
\caption{Example of a SWAP circuit that generates a wire permutation}
\label{fig:exsoa}
\end{figure}

This approach implies that $p_{out}$ and the SWAP circuit generating it are computed for more than a single CNOT (e.g. Fig.~\ref{fig:exsoa}). Both the set of remote CNOTs and the SWAP circuits are computed using heuristics (e.g. randomised algorithm in the IBM QISKit, A*-search \cite{zulehner2017efficient} or temporal planners \cite{venturelli2018compiling}).

\subsection{Moving on a (blue) CNOT circle}

Movement on a CNOT circle is equivalent to compiling CNOT gates sequentially. This is not to say that CNOT cannot be parallelised in the resulting $C'$ circuit. Parallelisation of a batch of CNOTs can be visualised on the torus: a CNOT is selected from the batch and compiled such that, for the remaining CNOTS, zero weights are placed on the edges connecting the configuration circles. Thus, for the first CNOT a kind of lookahead strategy \cite{wille2016look} has to be used to determine the configuration that will generate zero weight edges in the future.

Without discussing lookahead methods, compilation implies finding a \emph{good} configuration and then advancing on the CNOT circle. Thus, compilation is preceded by movements along the configuration circle whenever SWAP networks are used to prepare the configuration. But because remote CNOTs can be implemented also without SWAP networks, compilation of remote CNOTs can also have a different cost.

\subsection{Backtracking for TSP}

\begin{figure}[h]
\centering
\includegraphics[width=0.9\columnwidth]{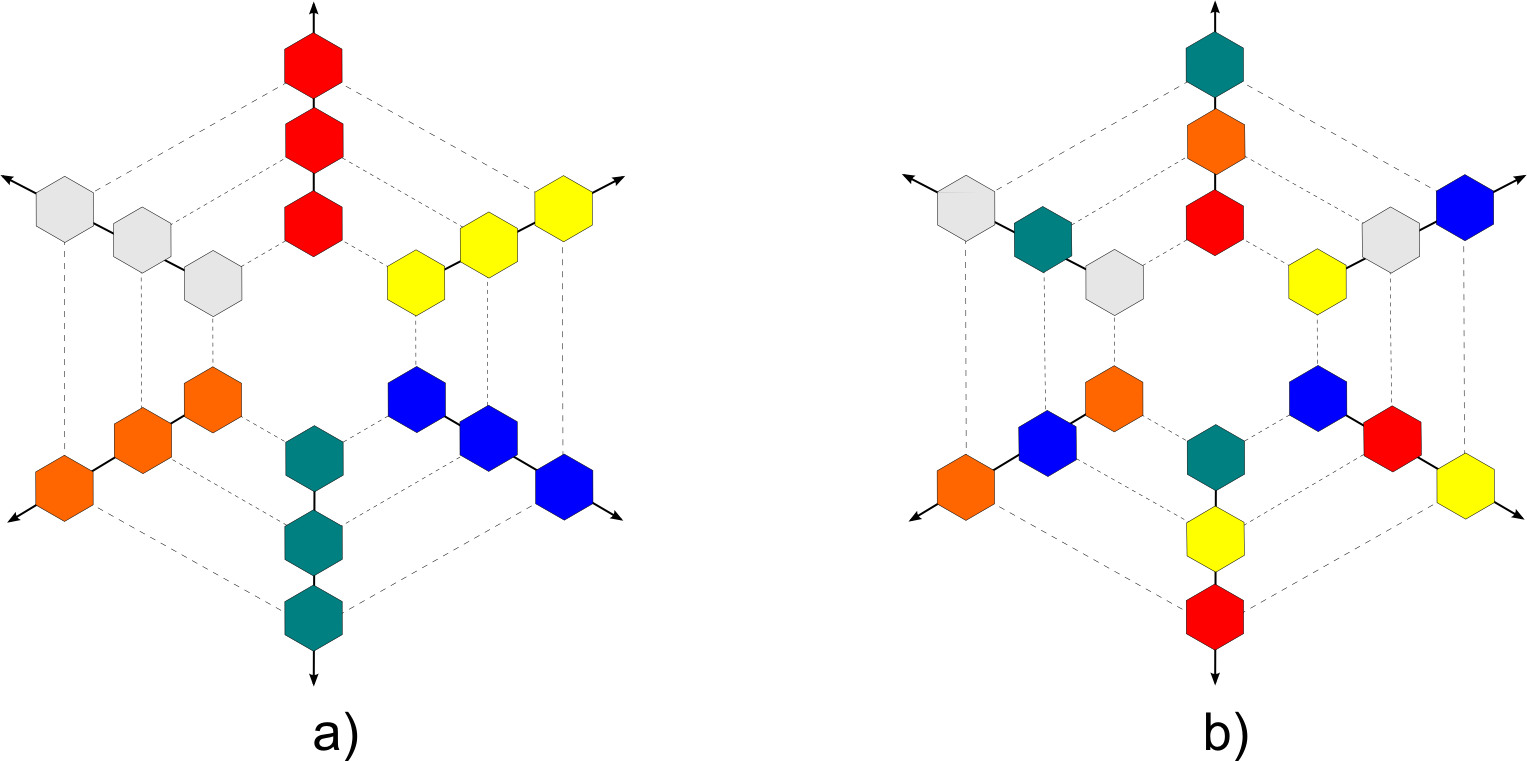}
\caption{Arrangement of configurations after MIM (a) and MI (b) swaps. Each configuration is coloured distinctively. MIM maintains the same configuration, while MI does not.}
\label{fig:sdconfigs}
\end{figure}

Having paralleled QCC to TSP, we can formulate a naive backtracking algorithm for compilation. The first step of the algorithm is to determine an initial configuration: how circuit qubits are mapped (allocated) to the NISQ (Fig.~\ref{fig:sdmoves}a). Afterwards, the first edge of the CNOT circle starting from this configuration vertex is traversed by choosing a coupling graph edge where to execute the CNOT. A new configuration is reached by using the MI swap strategy (Fig.~\ref{fig:sdconfigs}). The next torus edge traversal is prepared by moving around the configurations circle (Fig.~\ref{fig:sdmoves}b) and landing in a new configuration.

The backtracking step consists of two sub-steps: traversing the current configuration circle,  followed by traversing the CNOT circle. The backtracking step undoes the last CNOT compilation and moves along the previous configuration circle. This is equivalent to selecting a different edge where to map the remote CNOT that was just undone.

A solution is found each time a vertex from the outmost CNOT circle, marked by \emph{\ldots Stop}, is touched. Each solution is stored, and the best one is selected after the backtracking algorithm finishes: when all the cycles and configurations were naively considered.

Similarly to \cite{tan2020optimal}, it is possible to further increase to generality of the backtracking procedure by considering gate commutations on the blue rings. Then for each combination of the supported gate commutations, the torus has to be regenerated and the QCC procedure will have to be repeated.

\begin{figure}[h]
\centering
\includegraphics[width=0.9\columnwidth]{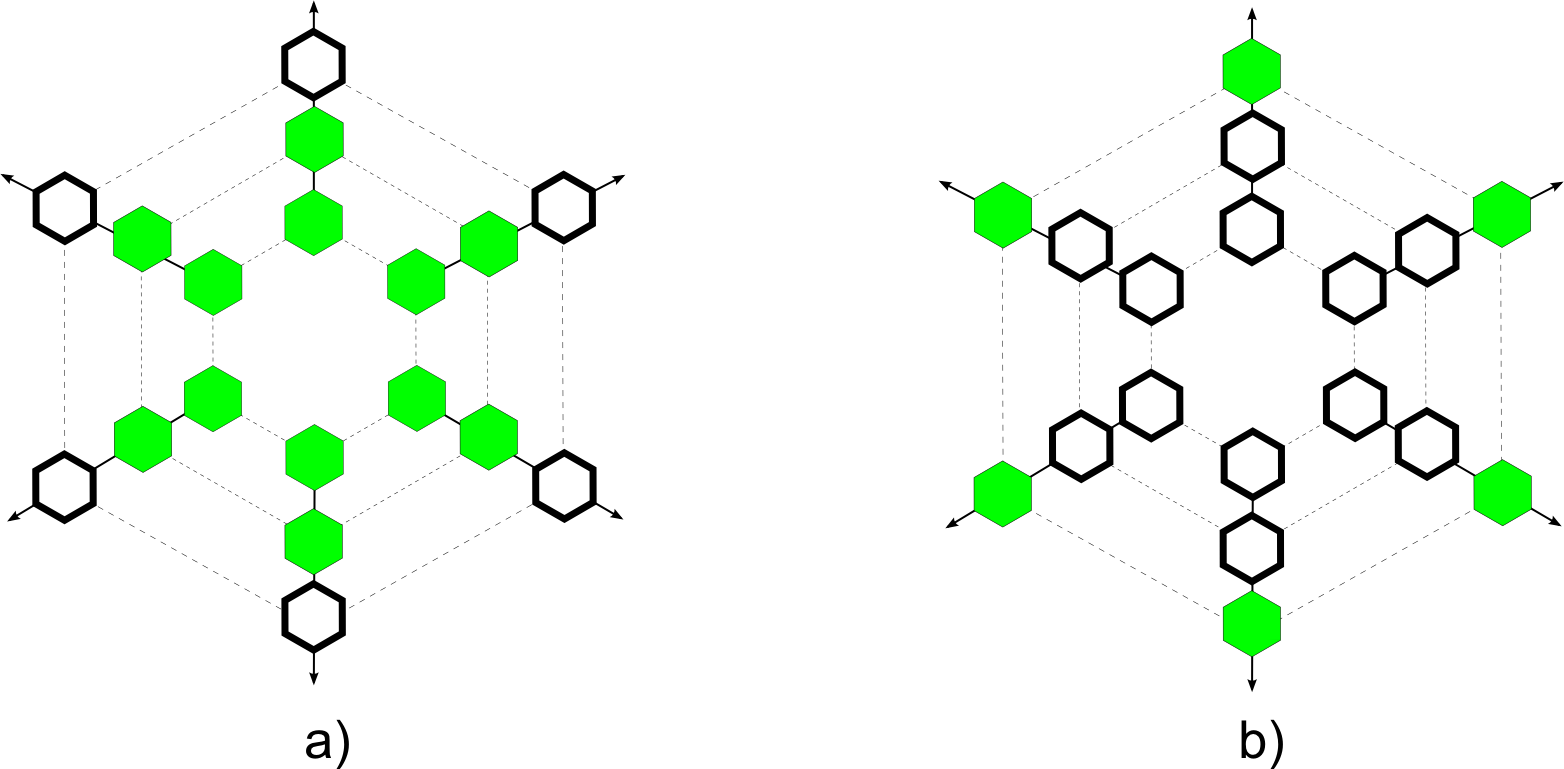}
\caption{The order of the concentric configuration circles is swapped after commuting CNOTs from $C$. In this example, there are two CNOTs to be scheduled: the first one is marked by green vertices, the second one by thick black stroked vertices (a single vertex of this CNOT is included in the figure). a) The green CNOT is compiled first, and the white one second; b) Assuming that the CNOTs can be commuted in the original circuit, the order of the vertices in each CNOT circle can be permuted. The white CNOT is compiled first and the green CNOT second (now, a single vertex of this CNOT is included in the figure).}
\label{fig:sdchains}
\end{figure}

\subsection{Pre- and post-processing}
\label{sec:heur}

The problem statement of QCC does not mention if $C$ is expressed using the universal gate set supported by the NISQ. If this is not the case, $C$ has to be translated to a functionally equivalent $C''$ that uses gates compatible with the NISQ gate set. This is a complex QCC pre-processing task with regard to the optimal number of resulting gates (e.g. \cite{shende2009cnot}), and does not fall within the scope of this work. Also, quantum algorithm and quantum hardware optimisations (cf. \cite{maslov2017basic}) are not considered parts of the general QCC framework.

The very high complexity of the exact method is a motivation for heuristics. It is useful to attempt to identify heuristic types and functionalities. As mentioned in Sec.~\ref{sec:intro}, compilation is the process of transforming a circuit $C$ into another circuit $C'$ that conforms to a set of constraints encoded into a coupling graph. Therefore, it is possible to \emph{preprocess} $C$ and \emph{postprocess} $C'$.

Preprocessing adapts $C$ for compilation, and it is viable to try and reduce the number of single qubit gates and CNOT gates by using, for example, template based optimisations \cite{saeedi2011synthesis}. Postprocessing can be template based too, as well as include recompilation of subcircuits of $C'$. For example, the IBM Qiskit uses this approach for single qubit gates, and this procedure was used by \cite{zulehner2019compiling}.

Heuristics can be included also for the previously discussed mapping problems. Selecting the start configuration (or any other configuration along the concentric cycles) could be performed using existing LNN optimisation methods, but cost models adapted to MI swaps should be formulated and analysed first. Another possibility is to collect all configurations generated along a CNOT-chain and try them out as start configurations. However, given the dimension of each configuration cycle, the collected configurations may be as good/bad as the initial one. Ranking coupling graph nodes is another heuristic for building the initial configuration \cite{ferrari2018demonstration}. The circuit mapping strategy presented in \cite{maslov2008quantum} would also fall in this category.

Traversal of edges along CNOT circles could be sped up by reducing the number of backtracking steps (minimum is zero), and to select from a few best possible edges for the mapping. The procedure for selecting \emph{the best coupling graph edge} is the following: 1) Shortest paths between all pairs of coupling graph vertices are computed using the Floyd-Warshall algorithm; 2) It is possible to add weights to the coupling graph edges (e.g. to prefer certain areas of the graph), or to treat the coupling graph as undirected; 3) Once a remote CNOT needs to be mapped to an edge, the sum of the distances between the coupling graph vertices where the qubits are located and each graph edge vertices is computed (e.g. Fig.~\ref{fig:heuristic}). The edge with the minimum distance sum is chosen, and, if multiple edges have the same distances, the last one in the list is chosen. Thus, the weighting function used for the coupling graph edges influences the edge selection. 

Edge mapping could be performed for multiple remote CNOTs in parallel, too. This possibility shows that the algorithm from \cite{zulehner2017efficient} is a heuristic fitting in the framework of this work.

\end{document}